\documentclass[epj,referee]{svjour}
\usepackage{graphics}
\usepackage{epsfig} 
\usepackage{amsmath}
\begin{document}
\title{Entanglement swapping between electromagnetic field modes and matter qubits}
\author{M. Kurpas \and E. Zipper %\inst{1}
}                     % Do not remove
\offprints{}  % Insert a name or remove this line
\institute{Institute of Physics, University of Silesia, ul. Uniwersytecka 4, 40-007 Katowice, Poland,\\ \email{mkurpas@us.edu.pl}}

\date{Received: date / Revised version: date}
\abstract{
Scalable quantum networks require the capability to create, store and distribute entanglement among distant nodes (atoms, trapped ions, charge and spin qubits built on quantum dots, etc.) by means of photonic channels.
We show how the entanglement between qubits and electromagnetic field modes allows generation of entangled states of remotely located qubits. We present analytical calculations of linear entropy and the density matrix for the entangled qubits for the system described by the Jaynes-Cummings model. We also discuss the influence of decoherence. The presented scheme is able to drive an initially separable state of two qubits into an highly entangled state suitable for quantum information processing.
\PACS{
      {03.67.-a}{Quantum Information}   \and
     {03.67.Bg}{Entanglement production and manipulation} \and
      {42.50.Pq}{Cavity quantum electrodynamics}
     } % end of PACS codes
} %end of abstract
\maketitle
\section{Introduction}
\label{intro}
\begin{figure}[h]
\begin{center}
\includegraphics[scale=0.4]{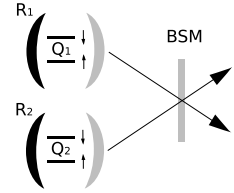} 
\end{center}
\caption{Sketch of the considered system. Each qubit $Q$ is placed in its own cavity $R$. The BSM performed on photons leaving the cavities entangles the qubits.}
\label{scheme}
\end{figure}

Entanglement being a quantum correlation between various parts of a system is required for quantum information processing. The quantum logic gates with qubits interacting directly with short range interaction are not suitable for linking distant nodes.
Quantum networks should be linked with light \cite{eck} which is the best long-distant carrier of information. Some schemes to entangle spatially separated, not directly interacting, pairs of qubits via single photon interference effects has been proposed \cite{zukow,moe,kok}.
In this article we perform analytical calculations of entanglement of two distant qubits by swapping \cite{zukow}.
To this end we consider a model of two quantum two-level systems each independently coupled to a single mode boson field (see Fig.\ref{scheme}). We operate in the regime in which the qubit-field coupling can be accurately described by the Jaynes-Cummings (J-C) Hamiltonian \cite{jaynes} giving two separate qubit-field entangled states. Then we subject the boson field mode from each pair to a joint measurement (BSM). This procedure projects two formerly independent qubits onto an entangled state that exhibits non-local quantum correlations.\\
In one of our recent papers we have discussed the entanglement of flux qubits using the above procedure \cite{zip}. As follows from detailed calculations, flux qubits are usually so strongly coupled to the electromagnetic field modes that the J-C model is not adequate and one has to perform numerical calculations or use the higher order approximations \cite{geller}. However there exists a range of qubits (atoms, ions \cite{haroche}, solid state charge qubits \cite{blais}) for which the inherent qubit-field coupling is weaker. We do not focus here on any of the specific examples but perform some general model calculations valid for systems which can be well described by the J-C Hamiltonian.
This model is exactly solvable and in this paper we take advantage of it and perform analytical calculations of some of the entanglement monotones. We derive the formulas for the linear entropy and show that it can be related in a simple way to the probabilities that can be easily measured \cite{stef}. 
In the first part of the paper all effects of dissipation and decoherence are assumed to be negligible over the studied time scales (strong coupling limit: $g>\kappa, \gamma$; $\gamma$, $\kappa$ are the decay rates of the qubit, field respectively). 
We then calculate the density matrix for the coherently coupled qubits and take into account the influence of the decohering environment.
\section{Conditional entanglement of qubits}
\label{ent}
Let us consider two separate qubit-field subsystems $(QR)_1$ and $(QR)_2$ described by the Jaynes-Cummings Hamiltonian 
\begin{eqnarray}
H_{(QR)_i} &=&\frac{\hbar \omega _{Q_i}}2\sigma _z+\hbar \omega _{R_i}\left( a^{\dagger
}a+\frac 12\right)-  \nonumber \\
&&\hbar g_i\left( a \sigma_{+}+a^{\dagger } \sigma_{-} \right)
\label{jc}
\end{eqnarray}
with the coupling constant $g_i$, field frequency $\omega_{R_i}$ and qubit frequency $\omega_{Q_i}$. Index $i=1,2$ numbers the subsystems.
We assume $g_{i} / \omega_{Q_i} < 0.1$. We have checked that in this regime the calculations with (\ref{jc}) are in agreement with exact numerical calculations \cite{zip,geller}.
The eigenstates of the uncoupled ($g_i=0$) Hamiltonian (\ref{jc}) are tensor products of the qubit and the field states $\vert \uparrow n \rangle =\vert \uparrow\rangle \otimes \vert n \rangle$ and $\vert \downarrow n \rangle=\vert \downarrow\rangle \otimes \vert n \rangle$; they describe the qubit in excited $\vert \downarrow \rangle$ and ground $\vert \uparrow \rangle$ states with a defined photon number $n$. The interaction term couples these states separating from the Hilbert space two-level subspaces $S_n \lbrace \vert \downarrow n \rangle, \vert \uparrow n+1 \rangle \rbrace$. 
Diagonalizing (\ref{jc}) at resonance  ( $\omega_{R_i} = \omega_{Q_i}\equiv \omega_R $) we obtain two eigenstates
\begin{eqnarray}
\vert +,n \rangle &=& \frac{1}{\sqrt{2}} \left( \vert \downarrow,n> + \vert \uparrow,n+1 \rangle \right), \\ \nonumber
\vert -,n \rangle &=& \frac{1}{\sqrt{2}} \left( \vert \downarrow,n> - \vert \uparrow,n+1 \rangle \right)
\end{eqnarray}
and corresponding energies
\begin{equation}
\frac{1}{\hbar}E_{\pm} = \left(n+1\right) \omega_R \pm g.
\end{equation}
In general $n$ can be an arbitrary integer number but the Hamiltonian (\ref{jc}) couples only $\vert n\rangle$ and $\vert n+1 \rangle$ states. For simplicity we assume $n=0$ that reduces $S_n$ to \\ $\lbrace\vert \uparrow 1 \rangle, \vert \downarrow 0 \rangle \rbrace$.
The entire system at $t=0$ is described by the vector
\begin{equation}
\vert \psi(0)\rangle=\vert \psi (0)\rangle_1\otimes \vert \psi (0)\rangle_2,
\label{psi_0}
\end{equation}
where $\vert \psi (0)\rangle_i$ describes the relevant $(QR)_i$ subsystem. \\
We discuss the entanglement for two different initial states:
\begin{equation}
\vert \psi (0)\rangle =\vert \downarrow 0 \rangle_1 \otimes \vert \uparrow1 \rangle _2 
\label{e0g1}
\end{equation}
and 
\begin{equation}
\vert \psi (0)\rangle  = \vert \downarrow 0 \rangle_1 \otimes \vert \downarrow 0 \rangle _2.
\label{e0e0}
\end{equation}
As the calculation procedure goes is the same way for both initial states we present below a detailed analysis only for the first case (\ref{e0g1}) giving merely the resulting formulas for the second one. \\
The unitary evolution of the $(QR)_i$s generated by (\ref{jc}) leads (for $g_i t\neq k \pi/2$, $k$ integer) to entanglement  of the qubit and field states
\begin{eqnarray}
\vert \psi (t)\rangle_1 &= &e^{-i\omega_{R_1} t} \left( \cos(g_1 t)\vert \downarrow 0\rangle - i \sin(g_1 t)\vert \uparrow 1\rangle \right), \nonumber \\
\vert  \psi (t)\rangle_2 &=& e^{-i\omega_{R_2} t} \left( -i \sin(g_2 t)\vert \downarrow 0\rangle +\cos(g_2 t)\vert \uparrow 1\rangle \right). 
\end{eqnarray}
During the evolution, the two $(QR)_i$ systems do not interact with each other and their state remains separable 
\begin{equation}
\vert \psi (t)\rangle=\vert \psi (t)\rangle_1\otimes \vert \psi (t)\rangle_2.
\label{psi_t}
\end{equation}
To entangle the qubits one needs to perform the BSM on the field modes $R_1$ and $R_2$ by projecting $\vert \psi (t)\rangle$ onto one of the Bell state and taking trace over the photonic degrees of freedom:
\begin{equation}
\vert QQ \rangle = Tr_R \left( \vert \psi^{-} \rangle_{RR} \langle \psi^{-} \vert \psi (t)\rangle \right) 
\end{equation}
To construct the projector we have chosen 
\begin{displaymath}
\vert \psi^{-} \rangle_R= 1/\sqrt{2} \left( \vert 01 \rangle -\vert 10 \rangle \right)
\end{displaymath} 
state, because of the easiness of its experimental verification. Then the resulting qubit-qubit state reads
\begin{eqnarray}
\vert QQ \rangle &=& e^{-i(\omega_{R_1} + \omega_{R_2})t}[\cos(g_1 t)\cos(g_2 t) \vert \downarrow \uparrow \rangle  \nonumber \\
&&+ \sin(g_1 t) \sin(g_2 t) \vert \uparrow \downarrow \rangle ].
\label{qq}
\end{eqnarray}

After normalization the qubit-qubit density matrix $\rho_{QQ}$ is given by:
\begin{equation}
\rho_{QQ}= \frac{\vert QQ \rangle \langle QQ \vert}{Tr \left( \vert QQ\rangle \langle QQ \vert \right) }
\end{equation}
To quantify the strength of quantum qubit-field and qubit-qubit correlations we calculate the linear entropy $S_{L}$ \cite{linent}
\begin{equation}
S_{L_{AB}} = 1 - Tr\left[ \left(\rho_{red}\right)^2 \right]
\label{sl}
\end{equation}
where $\rho_{red}= Tr_{B}\left[\rho_{AB}\right]$ is the reduced density matrix. $S_L=0$ for disentangled states and reaches 0.5 for maximally entangled states.\\
Using this formula we obtain the qubit-qubit linear entropy $S_L$
\begin{equation}
S_{L} = 1 -\frac{ \cos^4(g_1 t)\cos^4(g_2 t) + \sin^4(g_1 t)\sin^4(g_2 t)}
{\left( Tr\left[ \rho_{QQ}\right]\right) ^2}
\end{equation}
where
\begin{equation}
Tr\left[ \rho_{QQ}\right]=\cos^2(g_1 t) \cos^2(g_2 t) + \sin^2(g_1 t)\sin^2(g_2 t).
\end{equation}
We are also in position to calculate the qubit-qubit density matrix $\rho_{QQ}$ which can be reconstructed in quantum tomography experiments \cite{stef}.  
\begin{equation}
\langle\downarrow \uparrow \vert \rho_{QQ} \vert\downarrow \uparrow \rangle  = P_{2} 
\end{equation}
\begin{equation}
P_2= \frac{\cos^2(g_1 t)\cos^2(g_2 t)}{\cos^2(g_1 t) \cos^2(g_2 t) + \sin^2(g_1 t)\sin^2(g_2 t)}
\end{equation}
\begin{equation}
\langle \uparrow\downarrow \vert \rho_{QQ} \vert \uparrow \downarrow \rangle = P_{3} 
\end{equation}
\begin{equation}
P_3 =\frac{\sin^2(g_1 t)\sin^2(g_2 t)}{\cos^2(g_1 t) \cos^2(g_2 t) + \sin^2(g_1 t)\sin^2(g_2 t)}
\end{equation} 
\begin{equation}
\langle \uparrow \downarrow \vert \rho_{QQ} \vert\downarrow \uparrow \rangle = \langle \downarrow \uparrow \vert \rho_{QQ} \vert\uparrow \downarrow \rangle = \frac{ \sin(2 g_1 t) \sin(2 g_2 t)}{2 Tr[\rho_{QQ}]}
\end{equation}
The remaining matrix elements are zero. The signature of entanglement are the non-diagonal matrix elements. 
Between the probabilities $P_i$, $i=1\div4$ and the linear entropy there exists a simple relation
\begin{equation}
S_{L}=2 P_{2}P_{3}
\label{sl_p}
\end{equation}
Because we are working with the J-C Hamiltonian and due to the projection onto
 $\vert \psi^{-}\rangle_R$ state only two \\ ($\vert \downarrow \uparrow \rangle$and $ \vert \uparrow \downarrow \rangle$) from the four qubit-qubit states have finite probabilities.

For the case the system starts from the initial state
 $\vert \psi (0)\rangle = \vert \downarrow 0 \rangle_1 \otimes \vert \downarrow 0 \rangle _2$  (or $\vert \psi (0)\rangle = \vert \uparrow 1 \rangle_1 \otimes \vert \uparrow 1 \rangle _2$) \\
and using the same procedure we obtain the following formula for the linear entropy
\begin{equation}
S_{L} = 1 - \frac{ \sin^4(g_1 t)\cos^4(g_2 t) + \cos^4(g_1 t)\sin^4(g_2 t)}
{\left( Tr\left[ \rho_{QQ}\right]\right) ^2},
\label{e0e0_Sl}
\end{equation}
\begin{equation}
Tr\left[ \rho_{QQ}\right]= \sin^2(g_1 t) \cos^2(g_2 t) + \cos^2(g_1 t)\sin^2(g_2 t) 
\end{equation}
The corresponding probabilities are given by 
\begin{equation}
P_{2}= \frac{\sin^2(g_1 t)\cos^2(g_2 t)}{\sin^2(g_1 t) \cos^2(g_2 t) + \cos^2(g_1 t)\sin^2(g_2 t) }
\label{e0e0_P2}
\end{equation}
\begin{equation}
P_{3}=\frac{\cos^2(g_1 t)\sin^2(g_2 t)}{\sin^2(g_1 t) \cos^2(g_2 t) + \cos^2(g_1 t)\sin^2(g_2 t) }
\label{e0e0_P3}
\end{equation}
and
\begin{equation}
\langle \uparrow \downarrow \vert \rho_{QQ} \vert\downarrow \uparrow \rangle = \langle \downarrow \uparrow \vert \rho_{QQ} \vert\uparrow \downarrow \rangle =- \frac{ \sin(2 g_1 t) \sin(2 g_2 t)}{2 Tr[\rho_{QQ}]}.
\end{equation}
Again $P_1=0$, $P_4=0$ and the relation (\ref{sl_p}) is also true.

\section{Results}
\label{results}
In the first part of this section we present results obtained from above formulas for coherent evolution of the $QR$ subsystems. The influence of dissipation is discussed in the second part. The presented results are for the resonant case $\omega_{Q_i} = \omega_{R_i}=\omega_R$. 
 
\begin{figure}[h]
\begin{center}
\includegraphics[width=8.8cm]{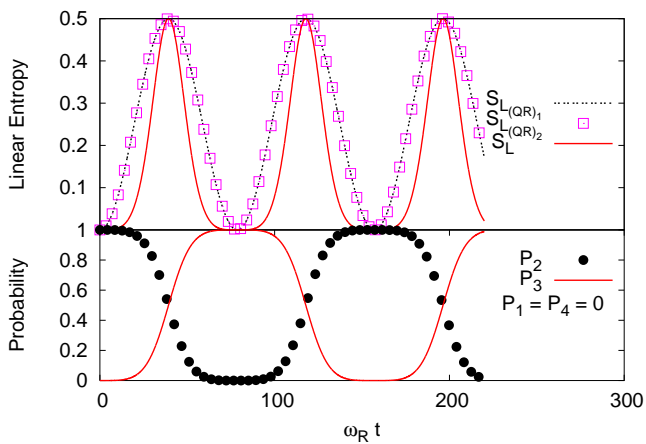}
\caption{(color online) Top: The linear entropies of qubit-field $S_{L_{(QR)_1}}$ (dotted line), $S_{L_{(QR)_1}}$ (open squares) and qubit-qubit $S_L$ (solid line). Bottom: The occupation probabilities of the qubit-qubit state after the BSM; $g_i/\omega_R=0.01$, $\omega_{Q_i}/\omega_{R} =1$.  }
\label{e0g1_lent}
\end{center}
\end{figure}

\begin{figure}[h]
\includegraphics[width=8.8cm]{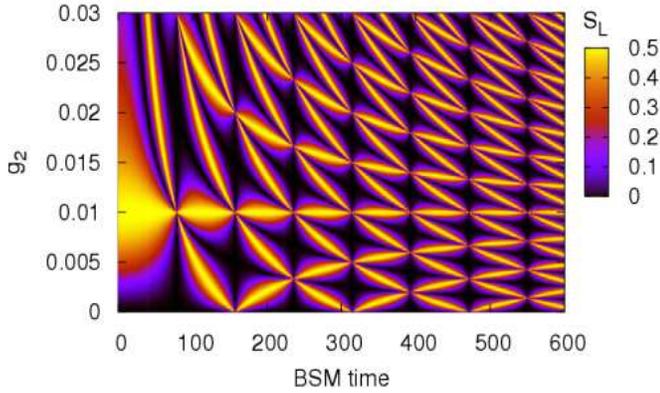}
\caption{(color online) The dependence of the qubit-qubit linear entropy $S_L$ on the BSM time for different values of the coupling constant $g_2$ for the initial state $\vert \downarrow 0 \downarrow 0\rangle$, $g_1/\omega_R=0.01$.  }
\label{e0e0_map}
\end{figure}

Let us first assume $g_1=g_2=g$; for concreteness we take e.g. $g/\omega_R=0.01$. 
Fig. \ref{e0g1_lent} depicts the results for the initial state given by (\ref{e0g1}).
At the top part we show the time evolution of the qubit-field linear entropy for $(QR)_1$
 ($S_{L_{(QR)_1}}$, dotted line) and $(QR)_2$ ($S_{L_{(QR)_2}}$, open squares) subsystems. The third curve ($S_{L}$, solid line) shows the qubit-qubit linear entropy as a function of time (henceforth called the 'BSM time') at which the BSM has been performed. It shows the strength of the $QQ$ correlations after the BSM performed at the moment
 $\omega_{R}t$. We see that the $QR$ entropies overlap and their shape exhibits oscillations. The value of $S_{L}$ is also oscillatory and depends on the values of $S_{L_{(QR)_i}}$ i.e. on the degree of entanglement of both subsystems. The maximal qubit-qubit correlations are obtained  by performing the measurement when the $QR$s are maximally entangled. The $S_L$ can be treated as a map of $QRs$ entanglement onto the $QQ$ entanglement: the more entangled are $QR$s the more entangled are qubits after the BSM. \\ 
The $QQ$ entanglement is also reflected in the occupation probabilities - it is presented in the bottom part of Fig. \ref{e0g1_lent}. The maximum of $S_{L}$, obtained for $gt=k\pi/4$, corresponds to equal probabilities of finding the qubits in $\vert \uparrow \downarrow \rangle$ and $\vert \downarrow \uparrow \rangle$ states i.e. the qubits are then in the Bell state. \\
The interesting situation arises for the initial state (\ref{e0e0}). As follows from Eqs (\ref{e0e0_Sl}-\ref{e0e0_P3}) for $g_1=g_2=g$ the formulas for $S_{L}, P_2$ and $P_3$ reduce to 
\begin{eqnarray}
S_{L}&=& \frac{1}{2} \left( \frac{\sin(2gt)}{\sin(2gt)}\right) ^4 = \frac{1}{2}\nonumber \\
P_2 =P_3 &=&\frac{1}{2} \left( \frac{\sin(2gt)}{\sin(2gt)}\right)^2 = \frac{1}{2},
\end{eqnarray}
i.e. after the BSM we always (except for $gt=k \pi/2$) get the qubits in the maximally entangled state. This is visible in Fig. \ref{e0e0_map} for $g_2=0.01$. For $gt=k \pi/2$ the vectors $\vert \Psi(t)\rangle_1$ and $\vert \Psi(t)\rangle_2$ represent separable states and the state vector of the whole system $\vert \psi(t)\rangle$ has no non-zero components along the direction of the Bell projector and thus the BSM is unsuccessful.

\begin{figure}[h]
\includegraphics[width=8.8cm]{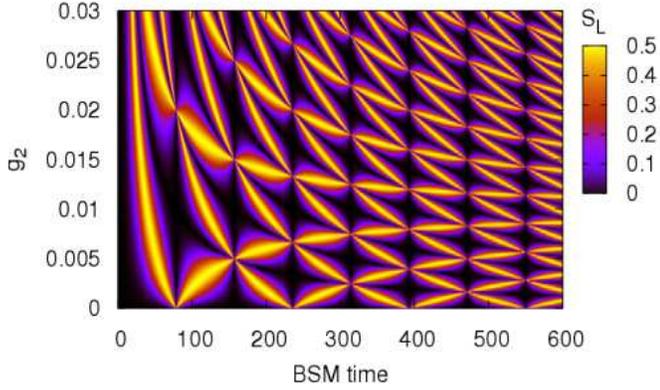}
\caption{(color online) The dependence of the qubit-qubit linear entropy $S_L$ on the BSM time for different values of the coupling constant $g_2$ for the initial state $\vert \downarrow 0 \uparrow 1\rangle$,  $g_1/\omega_R=0.01$. }
\label{e0g1_map}
\end{figure}

Next we investigate the behavior of the system for different values of $g_{i}$ in both $(QR)_i$ subsystems.
In Figs \ref{e0g1_map} and \ref{e0e0_map} we present the evolution of the linear entropy $S_{L}$ as a function of the BSM time for fixed $g_{1}/\omega_R=0.01$ and varying $g_2$ for the initial states $\vert \downarrow 0 \uparrow 1 \rangle$ and $\vert \downarrow 0 \downarrow 0 \rangle$ respectively. As one can see, these figures exhibit similar oscillations of entanglement in the almost whole range of parameters except for $g_1=g_2$. In this case in Fig. \ref{e0e0_map} the BSM results in the maximally entangled Bell state for every try of measurement for which there exists the non-zero component of $\vert \psi \rangle$ in the direction of the Bell projector. When $g_1$ is sufficiently far from $g_2$ the resulting entanglement is more or less regular function of the BSM time.

\begin{figure}[h]
         \includegraphics[width=8cm]{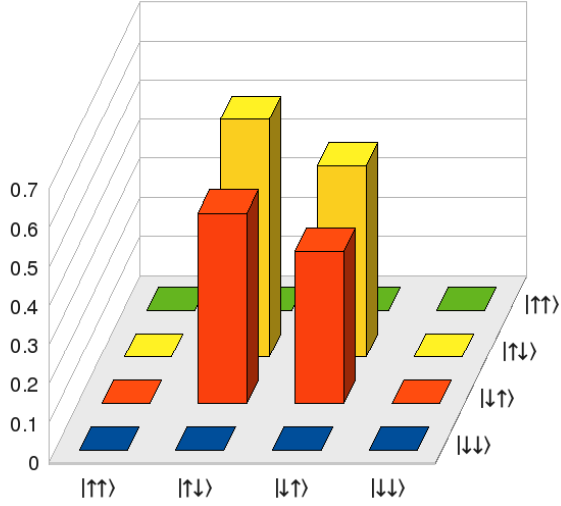}
  \caption{(color online) The qubit-qubit matrix elements at the BSM time $t=0.2 \mu s$ for coherent evolution of $QRs$. The initial state $\vert e0g1\rangle$, $\omega_{Q_i}=\omega_{R_i}= \omega_R$, $g_i/\omega_R=0.01, \omega_R=10GHz$. }
\label{e0g1_pure}
\end{figure}

\begin{figure}[h]
        \includegraphics[width=8cm]{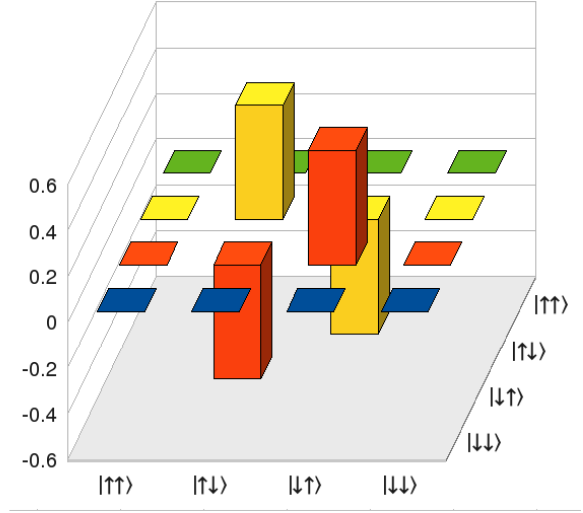}
         \caption{(color online) The qubit-qubit matrix elements at the BSM time $t=0.2 \mu s$ for coherent evolution of $QRs$. The initial state $\vert e0e0\rangle$, $\omega_{Q_i}=\omega_{R_i}= \omega_R$, $g_i/\omega_R=0.01$. }
\label{e0e0_pure}
\end{figure}

Now we discuss the density matrix $\rho_{QQ}$. Its elements are shown in Fig \ref{e0g1_pure} and Fig. \ref{e0e0_pure} for the BSM time $t=0.2 \mu s$ and for the initial states $\vert \downarrow 0 \uparrow 1 \rangle$ and $\vert \downarrow 0 \downarrow 0 \rangle$ respectively. These results convincingly show signatures of an entangled $QQ$ state namely the diagonal and non-zero off-diagonal matrix elements. We have chosen a relatively long BSM time so that the decoherence effects, which will be calculated below, are already visible. 
Till now we have presented calculations of the coherent evolution of the investigated system. Coupling to additional uncontrollable degrees of freedom leads to various decoherence processes in the qubit-field evolution which results in general in mixed states that can be described by the master equation in the Markov approximation. Following \cite{blais} we assume that the effect of environment can be included in terms of two independent Lindblad terms:
\begin{eqnarray}
\dot{\rho}_{QR}(t)=\left( L_H-\frac{1}{2}L_{\gamma}-\frac{1}{2}L_{\kappa}\right)  \rho_{QR}(t)
\end{eqnarray}
where the 'conservative part' is given by
\begin{eqnarray}
L_H(\cdot)=-i[H_{QR},\cdot]
\end{eqnarray}
whereas the 'Lindblad dissipators'
\begin{eqnarray}
L_{m}(\cdot)&=&A^\dagger_mA_m(\cdot) +(\cdot)A^\dagger_m A_m \nonumber \\ &-&2A_m(\cdot)A^\dagger_m
\end{eqnarray}
are expressed in terms of creation and annihilation operators 'weighted' by
suitable decoherence rates $A_{\gamma}=a\sqrt{\gamma}$ and $A_{\kappa}=\sigma_- \sqrt{\kappa}$, $m=\kappa,\gamma$. For concreteness we assume $\omega_R=10 GHz$, $\gamma=1MHz$, $\kappa = 3 MHz$.

\begin{figure}[h]
         \includegraphics[width=8cm]{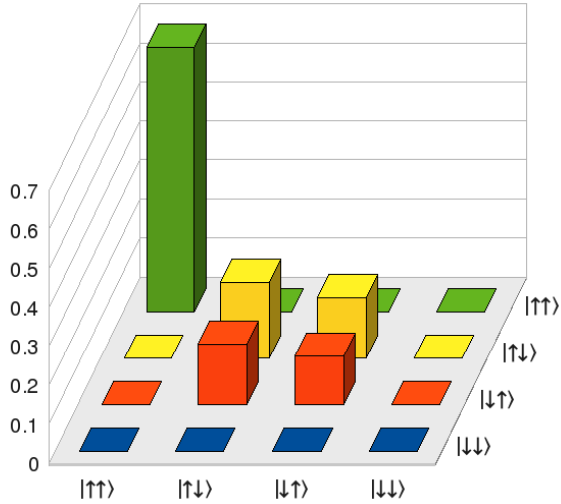}
     \caption{(color online) The qubit-qubit matrix elements at the BSM time $t=0.2 \mu s$ for dissipative evolution of $QRs$. The initial state $\vert e0g1\rangle$, $\omega_{Q_i}=\omega_{R_i}= \omega_R$, $g_i/\omega_R=0.01, \omega_R=10GHz$. }
\label{e0g1_dec}
\end{figure}

\begin{figure}
         \includegraphics[width=8cm]{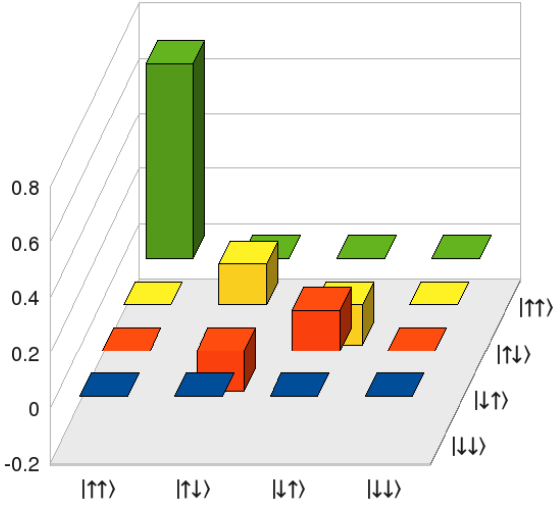}
         \caption{(color online) The qubit-qubit matrix elements at the BSM time $t=0.2 \mu s$ for dissipative evolution of $QRs$. The initial state $\vert e0e0\rangle$, $\omega_{Q_i}=\omega_{R_i}= \omega_R$, $g_i/\omega_R=0.01$. }
\label{e0e0_dec}
 \end{figure}

We have shown \cite{zip} that the BSM applied to the density operator of mixed states is a  well defined operation of projection and reduction which is completely positive and thus applicable to arbitrary density operator. Thus the whole procedure of swapping can be also performed if the decoherence effects are taken into account. \\
The results of the calculations of the matrix elements $\rho_{QQ}$ which follow from the solution of the master equation are presented in Fig.\ref{e0g1_dec} and Fig.\ref{e0e0_dec}. We see that dissipation causes significant decrease of the non-diagonal elements and the emergence of the additional matrix element $\langle \uparrow \uparrow\vert \rho_{QQ} \vert \uparrow \uparrow \rangle=P_4 \neq 0$ that means that the two qubits are already not in a pure state, but are still entangled.  

Dark counts in the detectors and imperfect mode matching of photons on the beam splitter can also reduce the fidelity of this scheme. However it has been shown \cite{kok} that this kind of entangling operations is rather robust against such errors.
Another important aspect of entanglement is the evolution of the state of correlated qubits. In the case discussed in this paper the $QQ$ state created by the BSM at certain moment $t$ does not evolve in time when neglecting decoherence. With decoherence taken into account the amplitudes of the diagonal and off-diagonal elements of $\rho_{QQ}$ decrease at the expense of the appearance of the qubits in the ground state.
In this paper we have not considered this problem in detail, concentrating mainly on the entanglement process. However, this problem has been studied in some papers (see e.g. \cite{rao}).
\section{Entanglement of spins encoded in quantum dots}
\label{dots}
Similar considerations can also be used to entangle qubits encoded in the electron spin of individual quantum dots as recently proposed in \cite{aws} (the spin states are very long lived with relaxation times of order of milliseconds).
\begin{figure}[h]
\includegraphics[width=8.8cm]{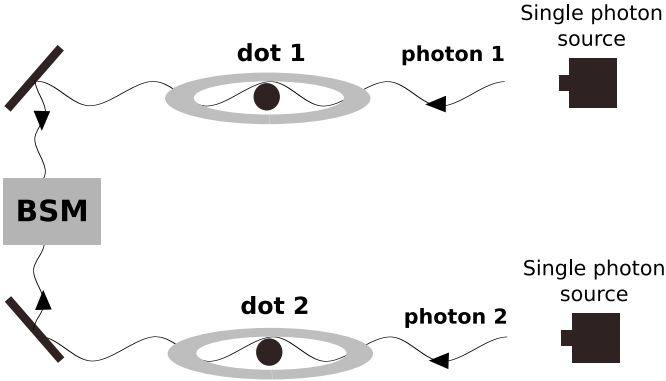}
\caption{Entanglement swapping procedure for photons and spins in quantum dots. The spin-photon interaction produces a conditional single-photon Faraday rotation. For the rotation angle $\pi/4$ the spin-photon state is maximally entangled. The BSM performed on such two photons produces entangled spins.}
\label{aws_scheme}
\end{figure}
In this approach (Fig. \ref{aws_scheme}) the establishment of spin-photon entanglement occurs through conditional (on the spin orientation) Faraday rotation of the incoming photon in a micro-cavity. Using the notation from \cite{aws} we obtain
\begin{equation}
\vert \Psi (T)\rangle = \vert \Psi_{phQ}(T)\rangle_1 \otimes \vert \Psi_{phQ}(T)\rangle_2,
\end{equation}
\begin{eqnarray} 
\vert \Psi_{phQ}(T)\rangle_1 &= &1/\sqrt{2}\left( \vert\searrow\hspace{-0.35cm}\nwarrow \rangle_1 \vert \uparrow\rangle_1 + \vert\nearrow\hspace{-0.35cm}\swarrow\rangle_1\vert\downarrow\rangle_1 \right)  \nonumber \\
\vert \Psi_{phQ}(T)\rangle_2 &= &1/\sqrt{2}\left( \vert\searrow\hspace{-0.35cm}\nwarrow\rangle_2 \vert \uparrow\rangle_2 + \vert\nearrow\hspace{-0.35cm}\swarrow\rangle_2\vert\downarrow\rangle_2 \right) ,
\end{eqnarray}
where $T$ is the interaction time between the electron spin and the photon in the micro-cavity,
$\vert\searrow\hspace{-0.35cm}\nwarrow \rangle$ and $\vert\nearrow\hspace{-0.35cm}\swarrow\rangle$ the photon states with a linear polarization rotated by $-\pi/4, +\pi/4$ respectively with respect to the state $\vert \leftrightarrow\rangle$ of linear polarization in the $x$ direction.
The BSM on the photons outgoing from the two micro-cavities conditionally leads to entangled qubit states
\begin{eqnarray}
\vert \Psi_{QQ} \rangle &=& Tr_{ph}\left( \vert \Psi^{-} \rangle_{phph} \langle  \Psi^{-}\vert \Psi\rangle\langle\Psi \vert \right)  \nonumber \\
& = &-1/ \sqrt{2}\left( \vert \uparrow\rangle_1 \vert \downarrow \rangle_2 - \vert \downarrow\rangle_1\vert \uparrow \rangle_2 \right) 
\end{eqnarray}
where 
\begin{equation}
\vert \Psi^{-} \rangle_{ph}= 1/\sqrt{2}\left( \vert\searrow\hspace{-0.35cm}\nwarrow \rangle_1 \vert\nearrow\hspace{-0.35cm}\swarrow\rangle_2 - \vert\nearrow\hspace{-0.35cm}\swarrow\rangle_1 \vert\searrow\hspace{-0.35cm}\nwarrow \rangle_2 \right)
\end{equation}
In this case the polarization degrees of freedom make up the photonic qubit.
This scheme provides a link between spintronic and photonic systems and can combine the advantages of both spintronic and photonic quantum information processing. 

\section{Conclusions} 
\label{conc}
Cavity quantum electrodynamics with individually addressable qubits (atoms, trapped ions, charge, flux or spin solid state qubits) is expected to provide a toolbox for quantum computing. The strong qubit-field coupling achievable in a high-finesse cavity can be accurately described by the Jaynes-Cummings model if $g/\omega_Q< 0.1$. This interaction is coherent permitting the transfer of quantum information between the qubit and the electromagnetic field modes. Photons are natural candidates as carriers of quantum information because they are highly coherent and can mediate interaction between different objects.\\
We have considered two remotely located stationary qubits each of which becomes entangled with the respective  electromagnetic field modes. Such qubit-field entanglement has been demonstrated e.g. for charge qubits \cite{blais} and for atoms \cite{haroche}. When the electromagnetic field modes are combined on a beam splitter their appropriate coincidence measurements ensure the entanglement of the two qubits. 
This procedure can be used both for pure and mixed qubit-field states and the results depend on the initial state of the system. The created entanglement survives in the presence of dissipation processes with the realistic decoherence rates of both qubits and fields.

The fact that the generation of the entangled state of qubits is conditional on the BSM on photons to some extent limits the application of this technique. However this is not an obstacle to what seems to be the most promising application - the entanglement of two distant qubits through the joint detection of the electromagnetic filed modes from two independent qubit-field subsystems. The above procedure can also be used to entangle qubits encoded in the electron spins of quantum dots \cite{aws}. 

The entangling operations discussed in this paper could be applied to generate cluster states of many qubits \cite{kok} which would be a step towards the distribution of entanglement through quantum networks \cite{pers}. 
\section*{Acknowledgments} 
Work supported by the Polish Ministry of Science and Higher Education under the grant N 202 131 32/3786.

\end{document}